# The Stoop-Squat-Index: a simple but powerful measure for quantifying lifting behavior

Stefan SCHMID[1,2]

[1]Bern University of Applied Sciences, Department of Health Professions, Division of Physiotherapy,
Spinal Movement Biomechanics Group, Bern, Switzerland
[2]University of Basel, Faculty of Medicine, Basel, Switzerland

**Correspondence:**
PD Dr. Stefan Schmid, Bern University of Applied Sciences, Department of Health Professions,
Murtenstrasse 10, 3008 Bern, Switzerland, +41 79 936 74 79, stefanschmid79@gmail.com

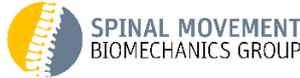

**ABSTRACT**

The widely held belief that squat lifting should be preferred over stoop lifting to prevent back injury is increasingly being challenged by recent biomechanical evidence. However, most of these studies only focus on very localized parameters such as lumbar spine flexion, while evaluations of whole-body lifting strategies are largely lacking. For this reason, a novel index, the Stoop-Squat-Index, was developed, which describes the proportion between trunk forward lean and lower extremity joint flexion, with possible values ranging from 0 (full squat lifting) to 100 (full stoop lifting). To enable the interpretation of the index in a real-life setting, normative values were established using motion capture data from 30 healthy pain-free individuals that were collected in the context of a previous study. The results showed mean index values of lower than 30 and higher than 90 for the most relevant phases of the squat and stoop movements, respectively, with mean index values differing significantly from each other for the full duration of the lifting phases. The main advantages of the index are that it is simple to calculate and can not only be derived from motion capture data but also from conventional video recordings, which enables large-scale in-field measurements with relatively low expenditure. When used in combination with lumbar spine flexion measurements, the index can contribute important information, which is necessary for comprehensively evaluating whole-body lifting strategies and to shed more light on the debate over the connection between lifting posture and back complaints.

## 1. INTRODUCTION

Most health care professionals and manual material handling advisors as well as guidelines issued by occupational safety organizations and even national health institutes such as the North American NIH or the British NHS promote the so called squat lifting technique as the "correct" and safe way compared to its opposite the stoop lifting technique, which is considered dangerous for the back and therefore strongly advised against (National Health Service [NHS], 2019; National Insitutes of Health [NIH], 2021; Nolan et al., 2018; Swiss National Accident Insurance Fund [SUVA], 2016). The squat lifting technique is thereby defined as flexing the knees and keeping the back as straight as possible (i.e. no forward flexion in the spine), while the stoop lifting technique is mainly achieved by a forward flexion of the spine without bending the knees.

However, despite these widely accepted guidelines, there is no consistent evidence which supports advocating squat over stoop lifting to prevent back injury. While some earlier observational evidence showed positive correlations between trunk forward lean and low back pain (LBP) incidence in occupational settings (Hoogendoorn et al., 2000), a recent meta-analysis revealed that greater lumbar spine flexion during lifting was neither a risk factor for LBP onset/persistence, nor a differentiator of individuals with and without LBP (Saraceni et al., 2020). These studies, however, only focused on partial aspects of lifting such as anterior trunk lean or lumbar spine flexion, and did not consider evaluations of whole-body lifting strategies, which might be important to shed more conclusive light on the debate over the connection between lifting posture and back complaints.

Such evaluations could be implemented for example by using laboratory-based three-dimensional optical motion capture techniques, which however would obviously not be suitable when aiming at large-scale measurements in occupational settings. A possible way for more easily determining whole-body movement strategies would be through an adequate index that could be derived for example from conventional video recordings. Currently available indices in the context of object lifting such as the well-known NIOSH Lifting Equation (Waters et al., 1993), however, do not consider any kinematic parameters and are therefore not suitable for this purpose.

For these reasons, this study aimed at introducing a novel index for quantifying the stoop-squat behavior, which can be easily derived from simple motion capture or conventional video recordings, and to establish normative values for healthy pain-free adults.

## 2. METHODS

### 2.1. Development of the Stoop-Squat-Index

To quantify whole-body strategies during object lifting, the Stoop-Squat-Index (*StSq*) was developed, which describes the proportion between trunk forward lean and lower extremity joint flexion based on the formula:

$$StSq = 100 - \left( \frac{(Vert\_HJC_{Standing} - Vert\_HJC_{Bending}) * 100}{Vert\_C7_{Standing} - Vert\_C7_{Bending}} \right) \qquad (1)$$

The variables *Vert_HJC* and *Vert_C7* represent the vertical positions of the hip joint center as well as the tip of the C7 spinous process, respectively, during standing and bending. An index of 0 thereby indicates a full squat movement, represented by a C7 downward displacement caused entirely by lower extremity joint flexion, whereas an index of 100 indicates a full stoop movement, represented by a C7 downward displacement caused entirely by trunk forward lean (Figure 1). Any value in between, e.g. an index of 50, indicates a lifting movement that is neither full squat nor full stoop, represented by a C7 downward displacement caused partially by lower extremity joint flexion and partially by trunk forward lean.

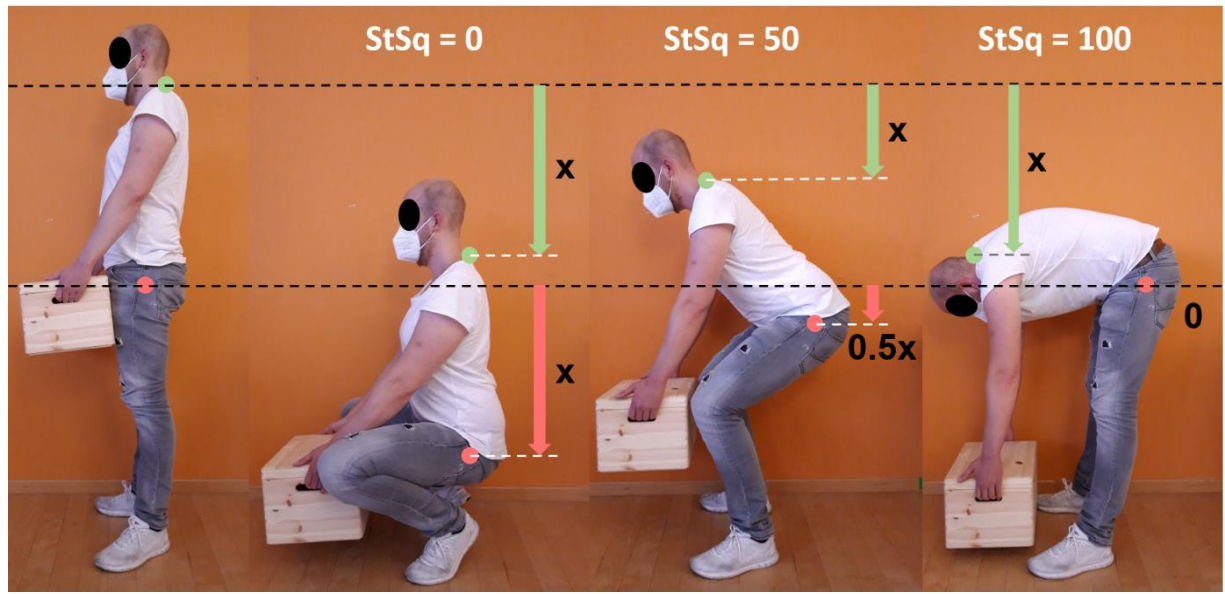

**Figure 1:** Interpretation of the Stoop-Squat-Index. An index of 0 indicates a full squat movement and an index of 100 a full stoop movement. Values in between indicate partially squat and partially stoop movements.

*2.2. Establishing normative values*

To enable the interpretation of the Stoop-Squat-Index in a real-life setting, normative values were established using motion capture data from 30 healthy pain-free individuals that were collected in the context of a previously conducted study. The protocol for this study was evaluated by the local ethics committee (Kantonale Ethikkommission Bern, Req-2020-00364) and all participants provided written informed consent. The sample consisted of 20 males and 10 females with a mean age of 31.8±8.5 years and a mean BMI of 23.3±2.4 kg/m$^2$. They were asked to perform five repetitions of each lifting up and putting down a 15kg-box using first a squat and then a stoop lifting technique. For squat lifting, participants were thereby instructed to lift the box with the back kept as straight as possible with mainly flexing the knees, and for stoop lifting, to lift the box by bending forward with a clear flexion in the spine and with the knees kept as straight as possible. Full-body kinematics were recorded using 58 retro-reflective skin markers (Schmid et al., 2017) and a 16-camera motion capture system (Vicon, Oxford, UK). Lifting up and putting down phases were identified using a previously described MATLAB-based event detection algorithm (R2020a, MathWorks Inc., Natrick, MA, USA) (Suter et al., 2020). To calculate the Stoop-Squat-Index, the vertical position of the tip of the C7 spinous process was derived directly from the marker placed over the C7 spinous process, whereas the vertical position of the hip joint center was approximated using the Plug-in Gait lower limb model (Davis et al., 1991) and the software Nexus (version 2.10.3; Vicon, Oxford, UK). For both movements, the index was calculated for the full duration of the lifting up and putting down phases, time-normalized to 101 data points and averaged over the five repetitions. Mean values and standard deviations were calculated to indicate typical values that can be expected in a healthy pain-free adult population.

To test the ability of the index for distinguishing between the squat and stoop movements, continuous index data were compared between the two movements using paired t-tests, implemented by the MATLAB-based software package for one-dimensional Statistical Parametric Mapping (SPM; spm1d-package, www.spm1d.org) (Pataky et al., 2013).

## 3. RESULTS

The calculations resulted in indices of lower than 50 and higher than 80 for the squat and stoop movements, respectively, during the first half of the lifting up and the second half of the putting down phases (Figure 2, top row). The mean indices for these phases were lower than 30 and higher than 90 for the squat and stoop movements, respectively. During the second half of the lifting up and the first half of the putting down phases, indices for the squat movement tended to increase towards 50, whereby indices for the stoop movement remained relatively consistent. It has to be noted that the indices of 2 participants during the lifting up and 1 participant during the putting down phase of the stoop movement dropped down to zero within the last and first 10%, respectively.

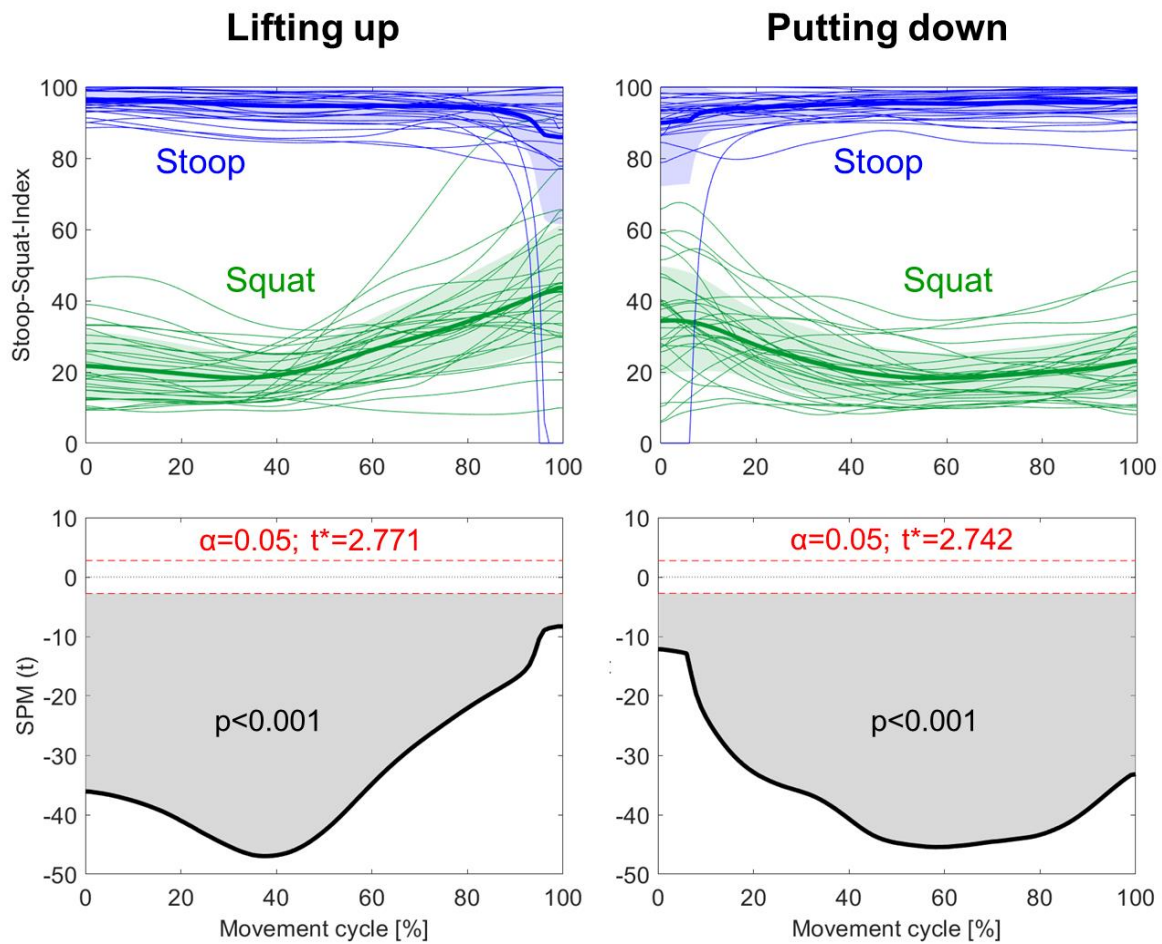

**Figure 2:** Top row: Normative values of the Stoop-Squat-Index for lifting up and putting down a 15kg-box based on a sample of 30 healthy pain-free adults, with the thick lines and shaded areas representing mean values und standard deviations. Bottom row: Comparison of the Stoop-Squat-Indices between the two lifting techniques using Statistical Parametric Mapping-based paired samples t-tests.

The comparisons of the indices between the squat and stoop movements revealed statistically significant differences ($p<0.001$) over the full duration of the lifting up and putting down phases (Figure 2, bottom row), indicating that the index was able to distinguish between the two movements at each instance of time, even during almost upright standing.

## 4. DISCUSSION

The current study aimed at introducing a novel and simple measure for quantifying the stoop-squat behavior during object lifting and to establish normative values for healthy pain-free adults. The proposed method, the Stoop-Squat-Index, describes the proportion between trunk forward lean and lower extremity joint flexion, with possible values ranging from 0 (full squat lifting) to 100 (full stoop lifting). Normative values showed mean values of lower than 30 and higher than 90 for the most relevant phases of the squat and stoop movements, respectively. Comparisons of the normative indices between the squat and stoop movements showed statistically significant differences over the full duration of the lifting phases.

The main advantages of the proposed method are that the index is simple to calculate and can be derived from uncalibrated (i.e. actual distances in meters unknown) conventional video recordings, enabling large-scale in-field measurements with relatively low expenditure. However, since the identification of the relevant landmarks (i.e. the tip of the C7 spinous process and the hip joint center) on the video images might not always be as clear as motion capture data, comparative evaluations are needed to determine the accuracy of such video-based measurements. Preliminary (unpublished) results of an ongoing validation study indicate that the measurement errors can be expected to be below 5 index points for the phases where the box is close to the floor during both a full squat and a full stoop movement. For the phases close to upright standing during a full stoop movement, on the other hand, errors might be somewhat higher. As shown in the results of the current study, indices of few participants dropped abruptly down to zero during these phases, which might attributable to slightly more flexed knee joints during the bending over phases resulting from hamstring tightness.

Due to a continuum of values that can be assigned for the movements between full squat and full stoop, this novel index is well suited for studies investigating possible causal effects between whole-body lifting strategies and LBP incidence as well as potential interactions between psychological factors and object lifting strategy. A recent study conducted by my research group and a group at the Balgrist University Hospital in Switzerland, for example, showed that a higher fear of "round-back lifting" (i.e. lifting with a flexed spine) was significantly associated with less lumbar spine flexion in healthy pain-free adults that were lifting a 5kg-box (Knechtle et al., 2021). However, since only lumbar spine flexion was evaluated, the study did not allow any conclusions on whether this reduced lumbar spine flexion also meant that the participants adopted more squat lifting behavior. For this reason, we decided to use the Stoop-Squat-Index developed in the current study to exploratively reanalyze the data and found that the fear of "round-back lifting" was actually not related to stoop-squat behavior (Schmid et al., 2021). These important findings clearly show that studies reporting altered lumbar spine flexion during lifting such as the ones summarized in the review of Saraceni et al. (2020) do not necessarily imply lifting strategy alterations on a whole-body level.

In contrast, it has to be kept in mind that the Stoop-Squat-Index does not allow any statements on how trunk forward lean is achieved, which could be either by tilting the pelvis anteriorly while keeping the spine straight or by flexing the spine while keeping the pelvis in an upright position. For this reason, it is suggested that the index is always used in combination with spinal flexion measurements. To avoid using sophisticated methods that require a laboratory setting, this could be achieved by using portable and easy-to-apply inertial measurement unit- or strain gauge-based systems such as the Epionics SPINE system, which has already been validated for the quantification of lumbar lordosis angles during object lifting (Suter et al., 2020).

In conclusion, the proposed index represents a novel and powerful measure for evaluating stoop-squat behavior during object lifting, which can fairly easily be derived from conventional video recordings with an expected high accuracy. When used in combination with lumbar spine flexion measurements, the index can contribute important information, which is necessary for comprehensively evaluating whole-body object lifting strategies.

## 5. CONFLICT OF INTEREST STATEMENT

The author declares no conflict of interest.

## 6. ACKNOWLEDGMENTS

The author thanks Christian Bangerter and PD Dr. Heiner Baur for reviewing and critically commenting on the first draft of the manuscript.

## 6. REFERENCES

Davis, R. B., Õunpuu, S., Tyburski, D., & Gage, J. R. (1991). A gait analysis data collection and reduction technique. *Human Movement Science*, *10*(5), 575–587. https://doi.org/10.1016/0167-9457(91)90046-Z

Hoogendoorn, W. E., Bongers, P. M., Vet, H. C. de, Douwes, M., Koes, B. W., Miedema, M. C., Ariëns, G. A., & Bouter, L. M. (2000). Flexion and rotation of the trunk and lifting at work are risk factors for low back pain: Results of a prospective cohort study. *Spine*, *25*(23), 3087–3092. https://doi.org/10.1097/00007632-200012010-00018

Knechtle, D., Schmid, S., Suter, M., Riner, F., Moschini, G., Senteler, M., Schweinhardt, P., & Meier, M. L. (2021). Fear-avoidance beliefs are associated with reduced lumbar spine flexion during object lifting in pain-free adults. *Pain*, *162*(6), 1621–1631. https://doi.org/10.1097/j.pain.0000000000002170

National Health Service. (2019). *Safe lifting tips*. https://www.nhs.uk/live-well/healthy-body/safe-lifting-tips/

National Insitutes of Health. (2021). *Back health: lifting with proper posture*. https://ors.od.nih.gov/sr/dohs/HealthAndWellness/Ergonomics/Pages/spine.aspx

Nolan, D., O'Sullivan, K., Stephenson, J., O'Sullivan, P., & Lucock, M. (2018). What do physiotherapists and manual handling advisors consider the safest lifting posture, and do back beliefs influence their choice? *Musculoskeletal Science & Practice*, *33*, 35–40. https://doi.org/10.1016/j.msksp.2017.10.010

Pataky, T. C., Robinson, M. A., & Vanrenterghem, J. (2013). Vector field statistical analysis of kinematic and force trajectories. *Journal of Biomechanics*, *46*(14), 2394–2401. https://doi.org/10.1016/j.jbiomech.2013.07.031

Saraceni, N., Kent, P., Ng, L., Campbell, A., Straker, L., & O'Sullivan, P. (2020). To flex or not to flex? Is there a relationship between lumbar spine flexion during lifting and low back pain? A systematic review with meta-analysis. *The Journal of Orthopaedic and Sports Physical Therapy*, *50*(3), 121–130. https://doi.org/10.2519/jospt.2020.9218

Schmid, S., Bangerter, C., Suter, M., & Meier, M. L. (2021). Fear-avoidance beliefs are not related to stoop-squat-behavior during object lifting in healthy pain-free adults. *Proceedings of the XXVIII Congress of the International Society of Biomechanics (ISB), Stockholm, Sweden*.

Schmid, S., Bruhin, B., Ignasiak, D., Romkes, J., Taylor, W. R., Ferguson, S. J., Brunner, R., & Lorenzetti, S. (2017). Spinal kinematics during gait in healthy individuals across different age groups. *Human Movement Science*, *54*, 73–81. https://doi.org/10.1016/j.humov.2017.04.001

Suter, M., Eichelberger, P., Frangi, J., Simonet, E., Baur, H., & Schmid, S. (2020). Measuring lumbar back motion during functional activities using a portable strain gauge sensor-based system: A comparative evaluation and reliability study. *Journal of Biomechanics*, *100*, 109593. https://doi.org/10.1016/j.jbiomech.2019.109593

Swiss National Accident Insurance Fund. (2016). *Hebe richtig - trage richtig*. https://www.suva.ch/de-CH/material/Sicherheitsregeln-Tipps/hebe-richtig---trage-richtig-44018d59315931

Waters, T. R., Putz-Anderson, V., Garg, A., & Fine, L. J. (1993). Revised niosh equation for the design and evaluation of manual lifting tasks. *Ergonomics*, *36*(7), 749–776. https://doi.org/10.1080/00140139308967940